\documentclass{article}
\usepackage{mrs2005,epsfig}
\begin{document} 
\title{Cosmological Simulations of Galaxy Formation I: Star Formation,
 Feedback, Resolution  and Matching the Tully--Fisher Relation
 (among other things).}

\author{Fabio Governato$^{1,2}$, Greg Stinson$^2$, J.Wadsley$^3$ \& T.Quinn$^2$}
\affil{$^1$INAF--Brera, Milano. ~$^2$University of Washington, Seattle. ~$^3$McMaster Univ., Hamilton.
[fabio,stinson]@astro.washington.edu}

\begin{abstract} 

We used fully cosmological, high resolution N-body+SPH simulations to
follow the formation of disk galaxies with a rotational velocity
between 140 and 280~Km/sec in a $\Lambda$CDM universe.  The simulations
include gas cooling, star formation (SF), the effects of a uniform UV
background and a physically motivated description of feedback from
supernovae (SN). Feedback parameters have been chosen to match the
star formation rate and interstellar medium (ISM) properties of local
galaxies. In cosmological simulations galaxies formed rotationally
supported disks with realistic exponential scale lengths and fall on
the I-band and baryonic Tully Fisher relations.  The combination of UV
background and SN feedback  drastically reduced the number of
visible satellites orbiting inside a Milky Way sized halo, bringing it
in fair agreement with observations.  Feedback delays SF in small
galaxies and more massive ones contain older stellar populations.
Here we focus on the SF and feedback implementations.  We also briefly
discuss how high mass and force resolution and a realistic
description of SF and feedback are important ingredients to match
the observed  properties of galaxies.

\end{abstract} 
 
\section{Introduction} 

N--Body/gasdynamical simulations have become the primary tools to
model galaxy formation in the hierarchical model of structure
formation.  They are necessary to follow the internal structure of
galaxies as well as the complex interplay between baryon
cooling/heating and energy injection from stellar processes.  In this
paper we describe a revised star formation algorithm[1] combined with
a physically motivated description of SN feedback first studied in
detail by Thacker \& Couchman[2].  We then discuss some results from
simulations of the formation of disk galaxies in a fully cosmological
$\Lambda$CDM context.  A complete description of the results presented
here will  follow in Stinson et al (S05) and Governato et al 2005
(G05).

SN winds or quenching of star formation by UV photons especially in
halos with small escape velocities have often[3,4] been mentioned as
important  in (a) shaping the angular momentum of disk
galaxies and the Tully - Fisher relation[5], (b) decreasing or stopping SF
in small galactic satellites[6] and (c) explaining the relation between
average stellar ages and total stellar masses in disk galaxies[7].
We studied the effect of feedback on the structure of disk galaxies
and their satellites by improving over previous works in a number of
ways: The mass and spatial resolution of these simulations (5-40
10$^5$ particles within the virial radius) are sufficient to (1)
resolve the structure of disks without being significantly limited by
resolution (2) to resolve the subhalo population for each galaxy in
our sample down to circular velocities of about 20\% of their parent
halo allowing us study the basic properties of galactic satellites.
We use the algorithm in cosmological simulations of individual high
resolution galaxies spanning a factor of 20 in mass and analyze their
properties. To better compare our models with observations we coupled
our outputs with GRASIL[8], to obtain the spectral energy
distribution (SED) of their stellar components that take into account
the effects of  metals and dust reprocessing.

\section{GASOLINE and the Star Formation/Feedback Algorithm} 
We used GASOLINE[9], a smooth particle hydrodynamics (SPH), parallel
treecode that is both spatially and temporally adaptive with
individual particles timesteps.  Simulations include the effect of i)
Compton and radiative cooling, ii) star formation and supernova
feedback and iii) a UV background following an updated version of
Haardt \& Madau ~[10]. Our algorithm can be broken down into three
main parts: (a) identifying the star forming regions, (b) forming
stars and (c) treating stellar evolution including such effects as
mass loss, SN winds and metal enrichment.

 {\bf The criteria for star formation} are as follow: {\it {\bf(a)}
Gas particle colder than T$_{max}$ = 30,000 K {\bf (b)}Local gas
density n$_{max}$ $>$ 0.1 cm$^{-3}$ {\bf(c)} Gas particle part of a
converging flow measured over the 32 nearest neighbors}.  We do not
include the criterion that gas particle be Jeans unstable as the
criterion proved to be resolution dependent (see S05). A minimum
overdensity  is also required to avoid spurious star
formation at very high z.

%
\begin{figure}  
\plottwo{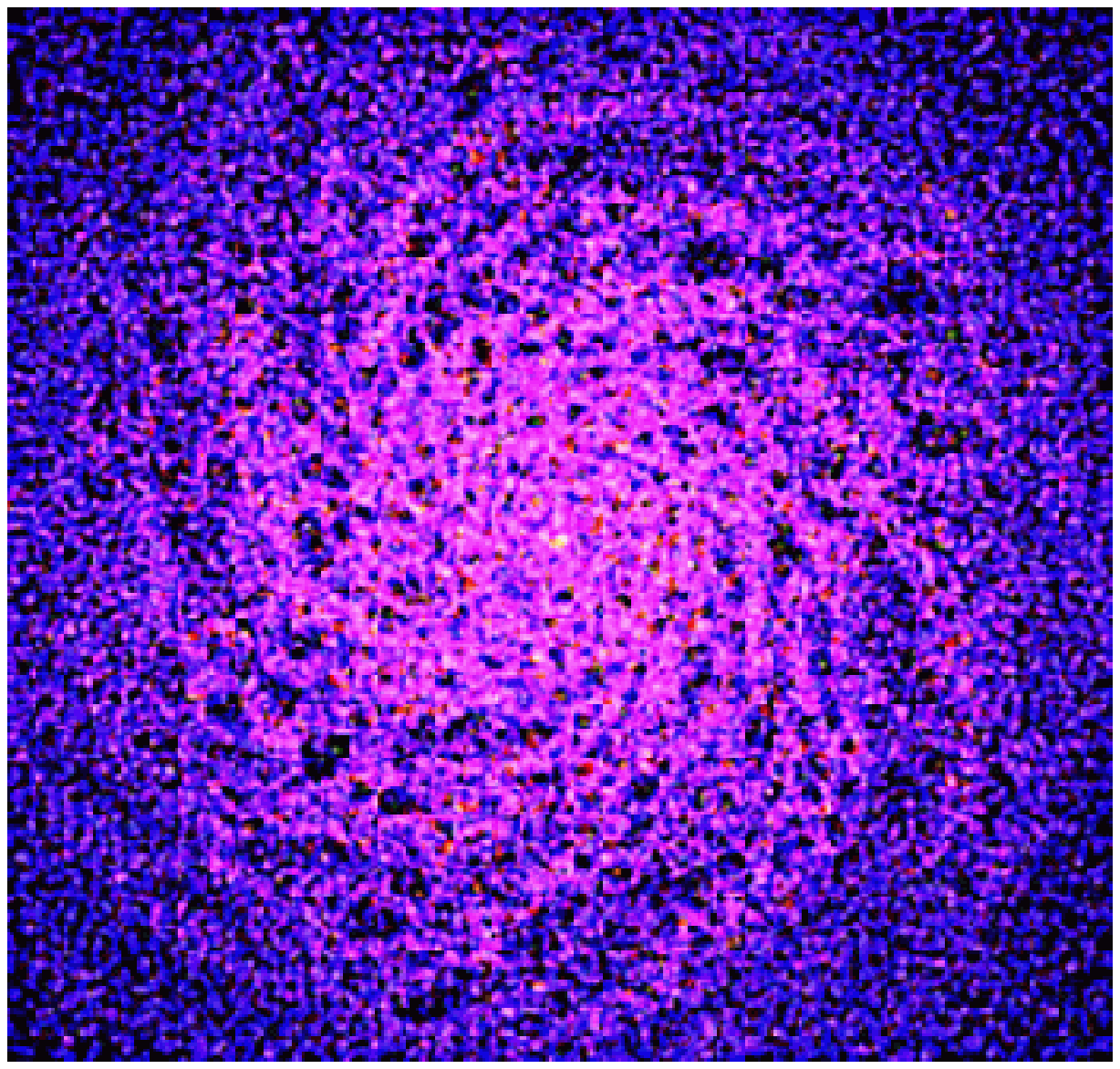}{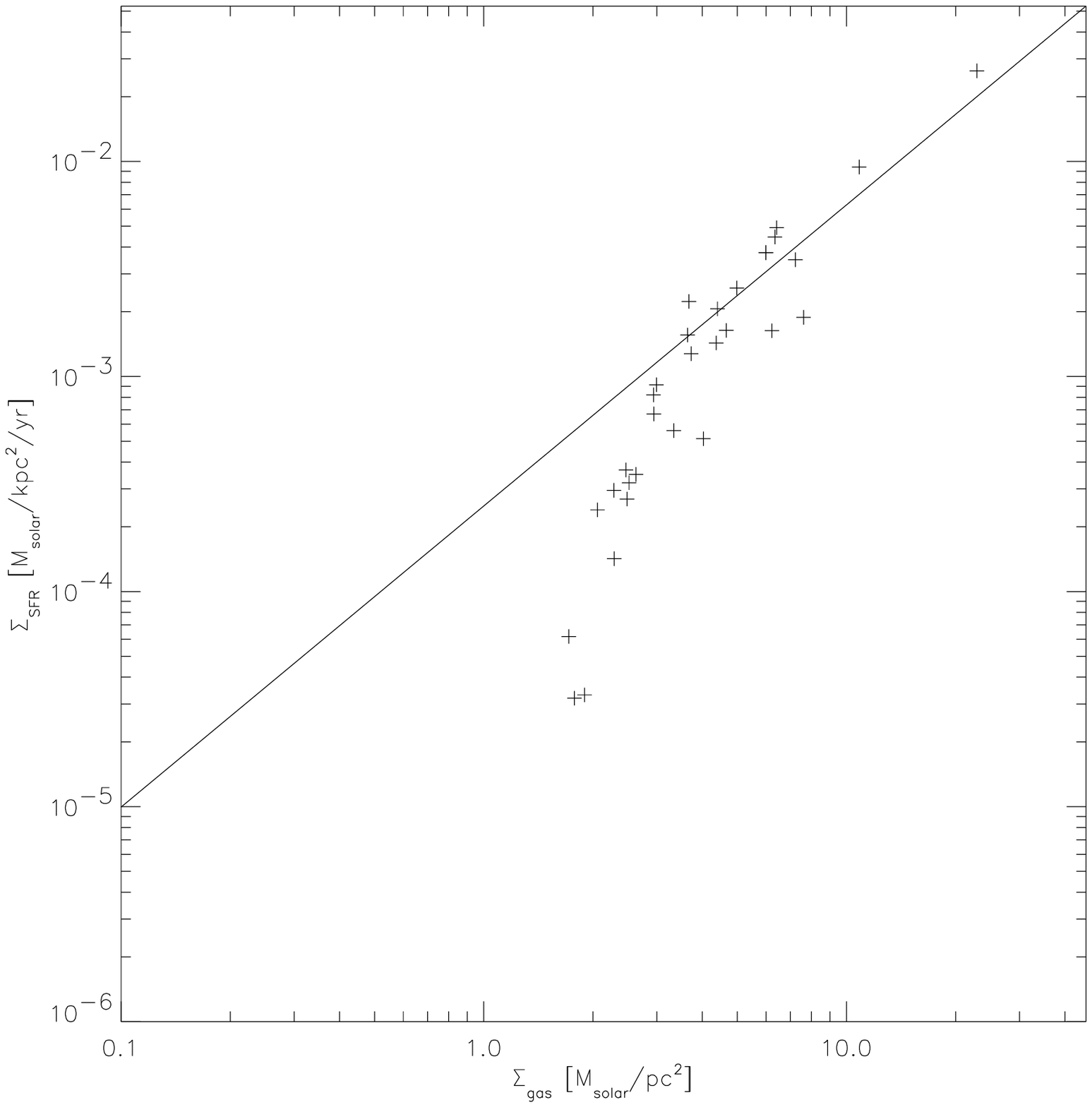}
\caption{Left panel: density map of the ISM of the isolated MW
model. Intra spiral arms holes are filled with low density hot gas
(not shown) that expands perpendicularly to the disk plane (see
Fig.2).  Right panel: SF rate vs local gas surface density in our
isolated MW model. Our model matches the slope and normalization of
Kennicutt's law~[11](solid line) 
}
\end{figure}

{\bf SF Efficiency:} we base the number of stars that form on the
proposed relation~[11] $\dot{\rho}_{SFR} \sim \rho_{gas}^{3/2}$, where
$\rho$ represents the volume density.  These formulations use the fact
that dynamical time, $t_{dyn} \sim \rho^{-1/2}$, so that

\begin{equation}
\frac{d\rho_{\star}}{dt} = c^{\star} \frac{\rho_{gas}}{t_{dyn}}
\end{equation}

where we have introduced a constant efficiency factor $c^{\star}$ that
 enables tuning of the star formation rate to observations We
 experimented with values in the range 0.01 -- 1. {\it and adopted a a
 value of $c^{\star}$ = 0.05}. Higher $c^{\star}$ values have been
 used when star forming regions are individually resolved[13]. The
 mass of star particles formed is fixed to 30\% of its parent gas
 initial mass. This choice is shown to give results weakly dependent
 on resolution.  Once the particle passes the above criteria, to
 implement eq.(1) in a discrete system we assign a probability $p$
 that a star will actually be spawned from its parent gas particle:

\begin{equation}
p < \frac{M_{SF}}{M_{GP}}(1-e^{-\frac{c^{\star} \Delta t}{t_{form}}})
\end{equation}

where we have introduced $M_{SF}$, the spawning mass for star
particles, $M_{GP}$, the mass of the gas particle that is creating the
star, $\Delta{t}$, the star formation timescale (1 Myr in all of the
simulations described in this paper) and $t_{form}$, which can be
either the dynamical time or the cooling time whichever is longer.
Gas particles in dense regions with shorter dynamical times will form
stars at a higher probability.

{\bf Feedback, mass loss and metal enrichment:} Once formed, a star
particle  can be considered as a single stellar population with
uniform metallicity.  We use a Miller -- Scalo IMF[14] (but we will
use Kroupa's IMF[15] in future work). The use of stellar lifetimes
and integration over the initial mass function provides the total mass
and number of stars that will explode during a timestep[16]. Metals
come from both SNs Ia \& II. The code also keeps track of mass loss
from stellar winds (e.g.  planetary nebulae). With the adopted IMF
star particles lose up to 30-40\% of their original mass as their
underlying stellar population ages. Neglecting this constant resupply
of gas would underestimate the SFR at late times. Feedback in our
simulations is purely thermal as we assume that the blastwaves of SNII
convert the kinetic energy of the ejecta into thermal energy on scales
smaller than those resolved in our simulations.

\begin{figure}
\begin{center}
\resizebox{17cm}{!}{\includegraphics{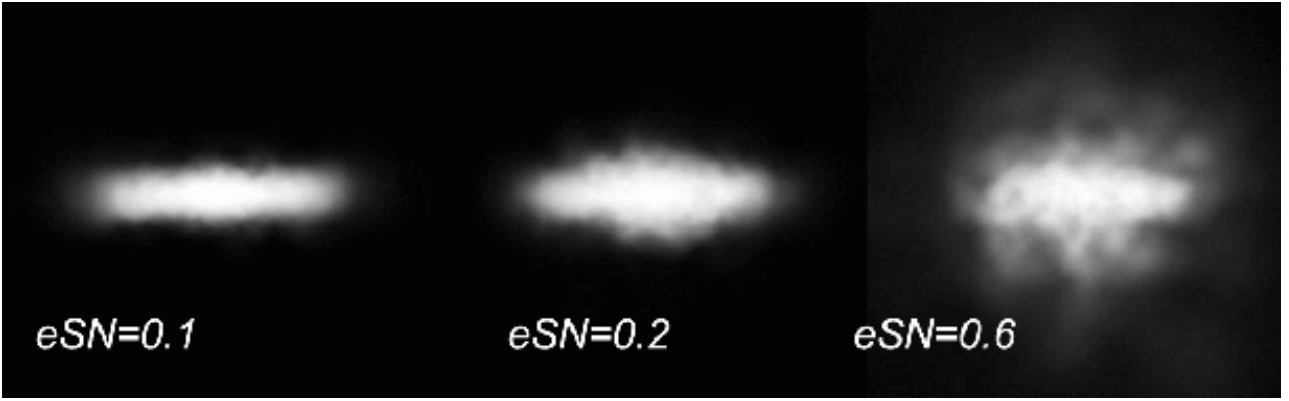}}\\
\end{center}
\caption{ The gas projected density for the isolated dwarf galaxy
model seen edge on as the SN efficiency {\it eSN} is increased. With
stronger feedback a galactic fountain is created. However gas loss
from the virial halo becomes important only in halos with circular
velocity V$_c$ $<$ 60 Km/sec.  Boxes are 20kpc across. }
\label{fig:feedback}
\end{figure}

Since some of the energy from SNs is quickly radiated away, we multiply
the number of SNII that explode by a fraction of the canonical
$10^{51}$ ergs/SN times an efficiency term (eSN), and distribute that
energy to the surrounding gas particles.  We explore efficiency values
in the range 0.1 to 0.6 {\it with eSN = 0.4 being our preferred
value.}  At an efficiency of 0.1, $7.65\times10^{47}$ ergs of energy
are deposited into the surrounding gas for every one $M_{\odot}$ of
star formed.  In this version of the feedback algorithm, energy and
metals are distributed using the smoothing kernel over the 32 nearest
neighbor particles (but see the Blastwave approach).  Our
implementation of feedback follows qualitatively the algorithm
implemented by Thacker \& Couchman[2]: we assume that the energy
released into the ISM turns into turbulent motions (at unresolved
scales) and stops the gas from cooling and forming stars. However, the
time scale for the cooling shutoff and the amount of mass affected are
now physically motivated and chosen following typical values from the
MacKee and Ostriker~[17] blast wave model (see next par.). In this version of the
feedback algorithm we {\it disable the radiative cooling for 20
million years} in a number of the nearest neighbor particles that
satisfy eq. (3):

\begin{equation}
\beta M_{SNII} > \frac{4 \pi r^3}{3} \rho_{ave}
\end{equation}

\noindent where $M_{SNII}$ is the mass of supernovae produced in a
star in a given timestep, r is the distance from the star to the gas
particle in question and $\rho_{ave}$ is the local gas density and
$\beta$ is a normalization factor. The maximum number of particles
that we can disable the cooling on is the number of SPH neighbors (32
in our simulations). For each of the simulations described in this work
we assumed a fixed value for {\it $\beta$ = 5000}.  We then performed
a large number of simulations to study the effect of different
combination of the main three star formation parameters: (a) the star
formation efficiency parameter c$^{\star}$, (b) the mass factor $\beta$
and (c) the SN efficiency {\it eSN} (Fig.1, left).  We applied our
algorithm to two N--body models of isolated galaxies having circular
velocities of 220 and 70 km/sec describing a Milky Way like and a
dwarf galaxy respectively.  They include a rotationally supported
stellar and gaseous disk, a bulge for the MW model and a
cosmologically motivated dark matter halo component.  Stars start with
a Toomre parameter Q=2. Disks are built to be stable to bar
instabilities. With the best parameters {\it indicated in italics}
model galaxies satisfy a number of observational criteria (see G05 for
more details): the amplitude of cold ISM turbulence, present day SFR
vs total stellar mass, R$_d$/R$_z$ (i.e disk radius/height scale
lengths ratios). The isolated MW model satisfies the Schmidt law and
its normalization (Fig.1, right). We then held the best parameters
fixed as we applied them to full fledged cosmological simulations.

{\bf The Blastwave Approach:} As described in S05, exploration of the
 $\beta$ parameter led us toward introducing an explicit blastwave
 solution[17].  This solution reduces the number of tunable parameters
 by providing both a maximum radius to which the blastwave explosion
 will reach and a time that the blastwave will keep the gas hot.
 $t_{max}$ corresponds to the end of the snowplow phase and $R_E$ to
 its radius and they are function of the local properties of energy
 injected, E$_{51}$, and the gas pressure $\tilde{P}$ and density
 n$_0$.

\begin{itemize}
\item $R_E$ (SN blast radius) = 10$^{1.74}E_{51}^{0.32}n_0^{-0.16}\tilde{P}_{04}^{-0.20}$ pc
\item $t_{max}$ (cooling shutoff timescale) = 10$^{6.85}E_{51}^{0.32}n_0^{0.34}\tilde{P}_{04}^{-0.70}$ years\end{itemize}

\begin{figure}[h]
\plottwo{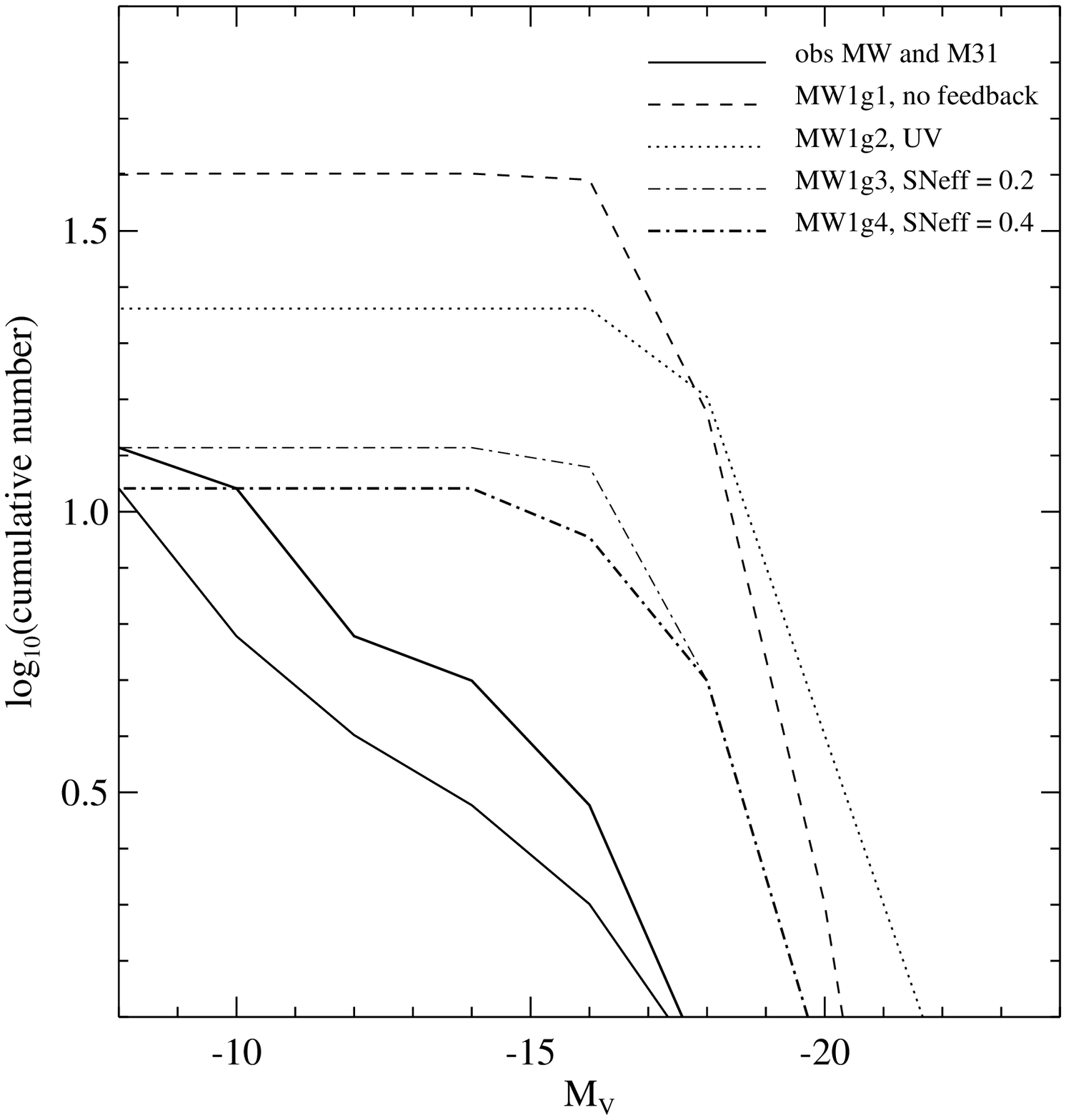}{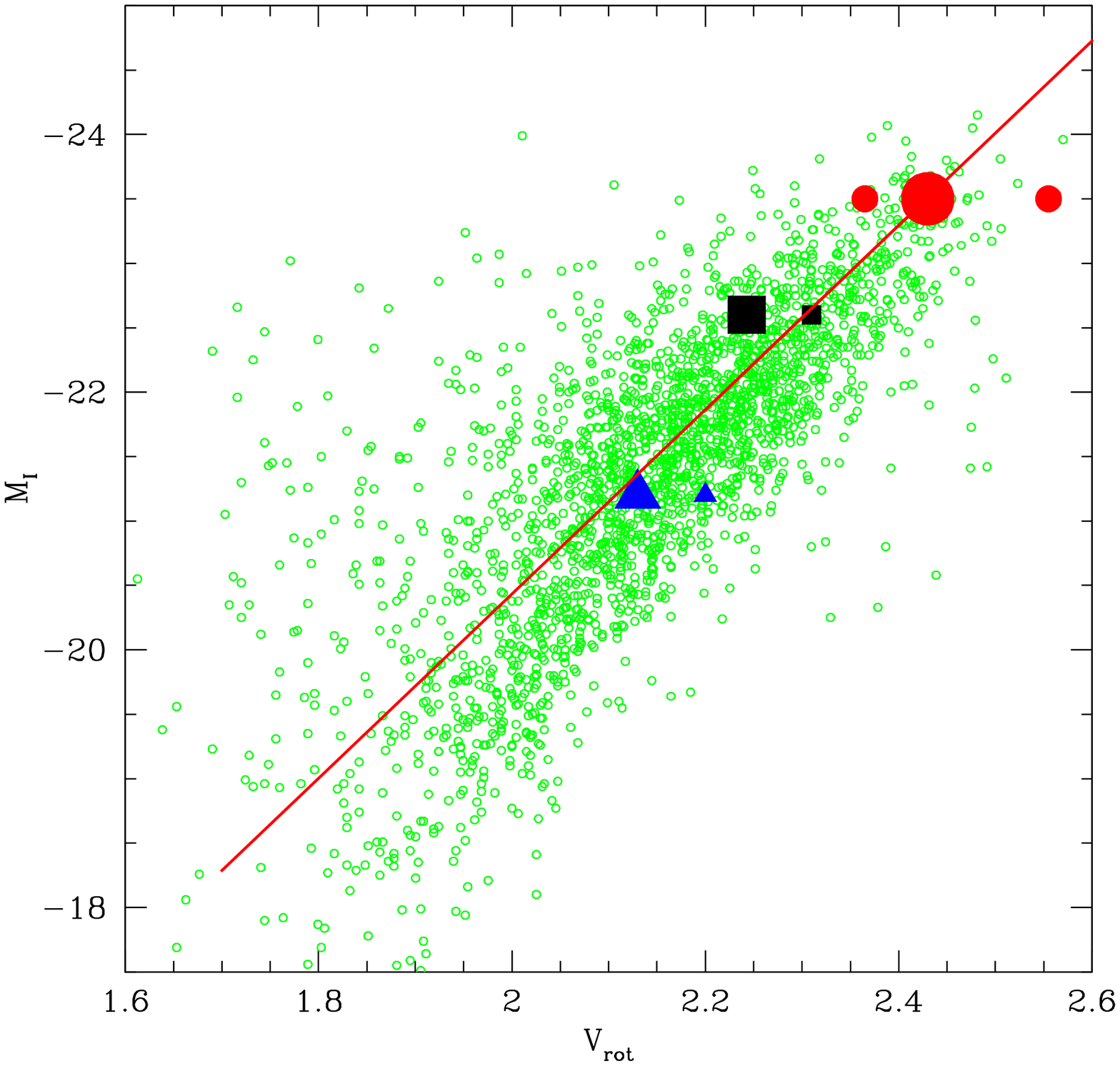}
\caption{Left panel:The luminosity functions of satellites within the
virial radius in all MW1 runs, compared with the luminosity functions
of both the Milky Way's and of M31's satellites.  The two runs without
supernova feedback produce far too many satellite.  However, the two
runs with increasing supernova feedback produce a number of satellites
that well matches that of the Milky Way and M31 satellite populations,
although the simulated satellites are still too bright. Right: The
Tully Fisher relation using a data compilation Giovanelli \&
Haynes. The red line is a fit to published data[19]. Solid triangle:
DWF1, Solid Square: MW1, Solid Dot: GAL1. Bigger dots shows V$_{rot}$
measured at 3.5R$_d$. Smaller dots shows the effect of measuring
V$_{rot}$ at 2.5R$_d$. The rightmost red dot uses the stellar peak
velocity}.
\label{twobarrel}
\end{figure}

With this approach, we only disable cooling in a fraction of the
particles within the smoothing length. We verified that results
presented here do not change significantly as the more sophisticated
feedback recipe is adopted. However, the ``blastwave approach''
further reduces star formation in halos with mass $< 10^{10} M\odot$.
This refined approach has as free parameters only c$_*$, eSN, wich we
contrained from local galaxies and will become our standard for future
works. While scales below the resolution (0.2-1kpc) remain unresolved,
the approach outlined above gives a simple, but physically motivated
description of star formation and feedback with no fudge factors {\it
at the resolved scales} that can be tested and ultimately
falsified. Additional levels of complexity can eventually be added.

\begin{table*}
\centering
\begin{tabular}{cclcccc}
\hline\hline
Run & Virial Mass  & 
Vc$^{a}$ & spin & 
Last major merger  &
  $\epsilon$ &
 N$_{tot}^{b}$ at z=0 \\
${}$ & M$\odot$ & Km/sec & $\lambda$ & z & kpc & dark+gas+stars \\
\hline
DWF1 &1.6 10$^{11}$  &   70  & 0.01  & 2.2  &  0.3  & $\sim$ 860.000 \\
MW1$^1$ & 1.15 10$^{12}$  &  134    & 0.04  & 2.5  & 0.6-0.3$^1$  & $\sim$ 700.000 - 4.000.000$^2$ \\
GAL1& 3.1 10$^{12}$ &  185    & 0.035 & 2.75  & 1.  & $\sim$ 480.000  \\

\hline
\end{tabular}
\caption[Summary of the properties of the three cosmological halos]
{Summary of the properties of the three cosmological halos: $^a$:
Circular velocity at virial radius, $^b$ number of gas and star
particles changes slightly depending on eSN. $^{1\&2}$ smaller
softening/larger N values are for the z=0.5 Super High Res.Run. }
\label{tbl-3}
\end{table*}

\section{Cosmological simulations}

For our cosmological runs we selected three halos from low resolution,
dark matter only simulations in a concordance flat $\Lambda$CDM
cosmology: $\Omega_0=0.3$, $\Lambda$=0.7, $h=0.7$,
$\sigma_8=0.9$. From a 28.5 and 100 Mpc box, low resolution
simulations, we selected and resimulated at higher resolution three
halos with mass between 1.6 $\times$ 10$^{11}$ and 3 $\times$
10$^{12}$ M$_{\odot}$ and a relatively quiet merger history: good
candidates to host spiral galaxies.  The ``zoomed in''[18] approach
for the initial conditions is the most realistic as torques from
distant objects and infall from surrounding regions are more correctly
described than in isolated initial conditions of uniform collapsing
regions with solid body rotation.  Table 1 shows the main parameters
of the simulations and of the selected halos, whose mass should
correspond to a small galaxy (DWF1), a Milky Way sized one (MW1) and a
more massive halo (GAL1) likely associated with early type spirals or
S0s. Halos were selected with somewhat early but not uncommon epochs
for their last major merger event {\it to test the null hypothesis
that the simulations should produce disk dominated galaxies}. The MW1
simulation was run with increasing feedback (SN efficiency 0.2,0.4 and
0.6) and once with 8 times more particles to test resolution effects
(to z = 0.5). This ``super'' run is the highest resolution run to date
of a MW sized galaxy with more than 4 million particles, a gas
particle mass of only 10$^5$ M$\odot$ and force resolution of 0.3 kpc.
By the present time all three galaxies show an extended disk component
supported by rotation.  In all but the most massive galaxy the bulge
component is sub-dominant, with the outer stellar halo contributing
only ~ 10\% to the total stellar mass. Once corrected for dust
reprocessing in their disks, the three galaxies show fairly blue
colors (B-V = 0.6).

The simulated galaxies share some important properties with real ones:
since feedback removes gas and reduces drastically star formation in
halos with V$_c$ ( ${\sqrt{M(r<R)/R}}$) $<$ 50 Km/sec, MW1 has only
about 10 satellites with a significant stellar component (Fig.3) with
another 20 dark satellites.  The smaller galaxy (DWF1) has all but one
satellite completely dark.

One difference with most previous works is that the galaxies fall
easily on the I-band Tully--Fisher relation[19] (Fig.3) and provide an
equally good match to the baryonic Tully Fisher[20]. Following the
observational sample for V$_{rot}$ we used the azimuthally averaged
rotation curves (using cold gas as a tracer) of galaxies at 3.5 and
2.2R$_d$, where R$_d$ is the disk scale length in the approximate I
band (stars younger than 5 Gyrs).

\begin{figure}[h]
\plottwo{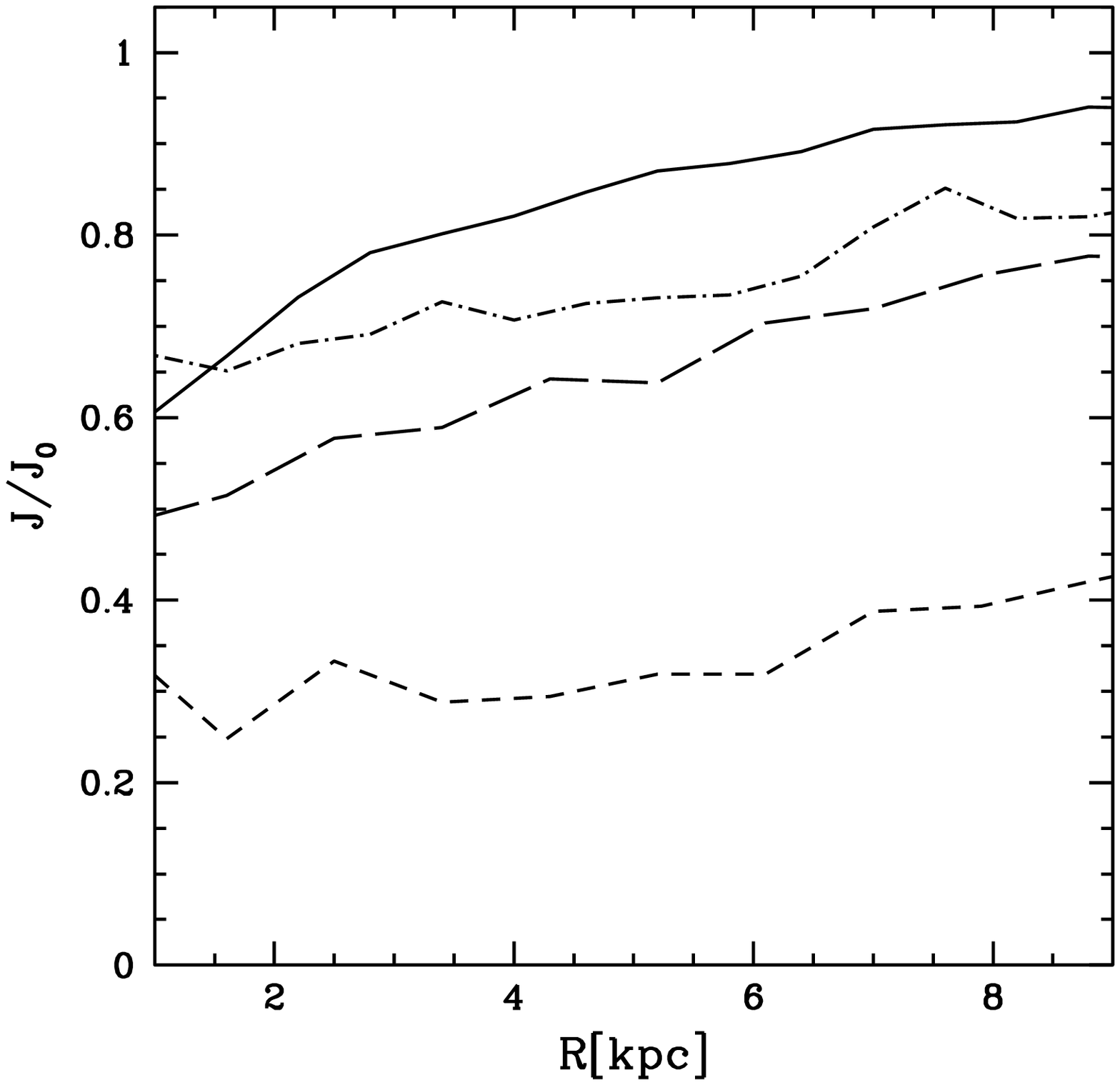}{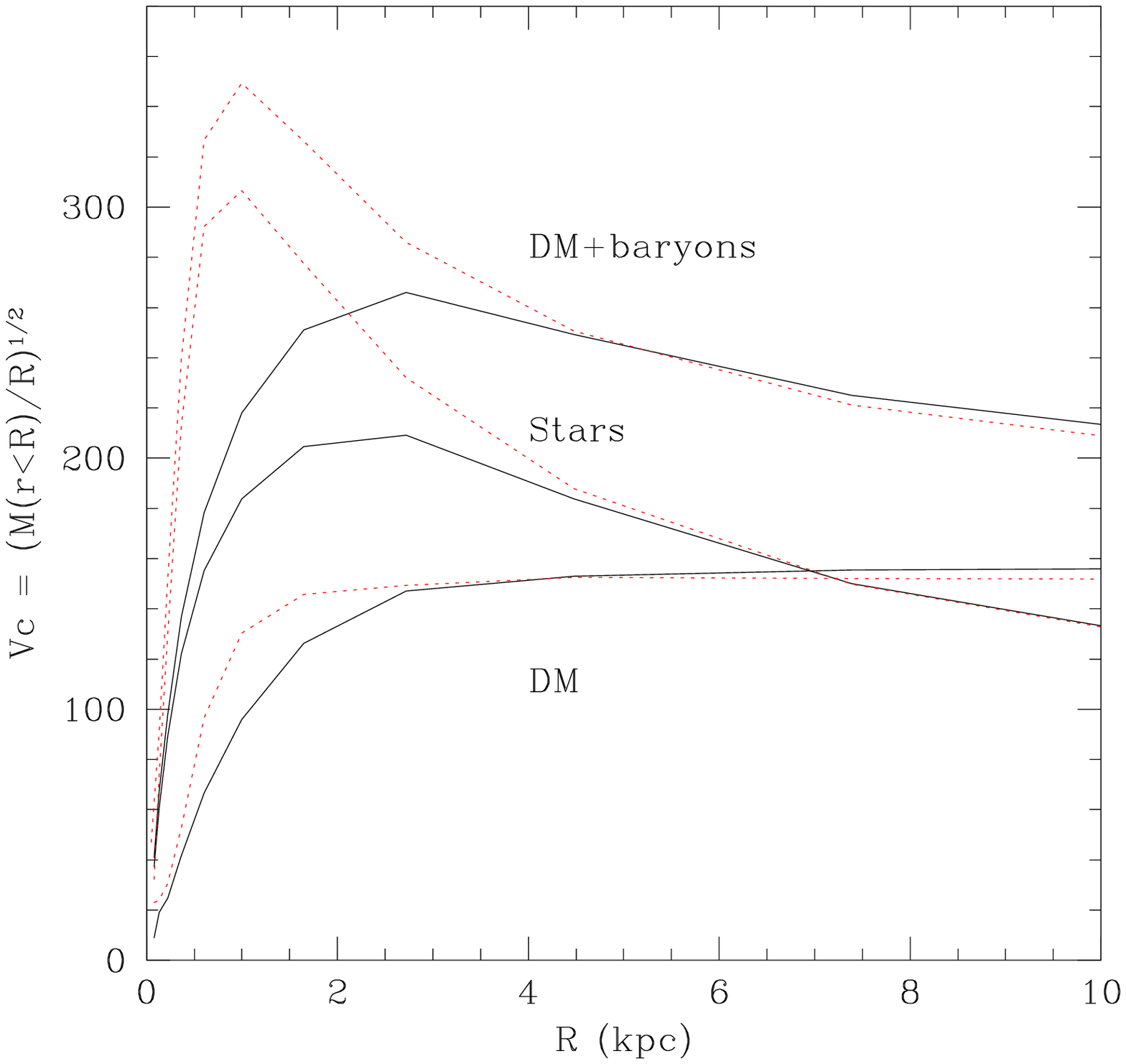}
\caption{Left panel:angular momentum loss in an isolated disk galaxy
 model[18]. Simulations were run for 6 Gyrs.  The y axis shows the
 fractional angular momentum loss for all the baryonic material in the
 disk as a function of radius. We verified that most of the ``lost''
 angular momentum is transferred to the DM component.  Continuous
 line:N$_{DM}$ = 100000, Nstar=200000, Ngas=5000. Dotted short dashed:
 N$_{DM}$ and Nstar reduced by a factor of five. Long dashed: Nstars
 reduced by a factor of 25. short dashed:N$_{DM}$= 4000, Nstar=8000,
 Ngas=1000. At the lowest resolution the disk undergoes the
 catastrophic angular momentum loss reported in some early
 simulations.  Right Panel: The circular velocity V$_c$ =
 ${\sqrt{M(r<R)/R}}$ for our MW1 halo at z=0.5. Halos have
 ~600.000/4.000.000 particle and force (spline) softening of 0.6--0.3
 kpc respectively (solid vs dotted). The effect of low resolution at
 scales $<$ 3-5 times the softening is evident with a much more
 pronounced bulge. In our simulations V$_c$ = V$_{rot}$, (the
 tangential velocity of cold gas \& stars) at $>$ ~R$_{disk}$.}
\label{twobarrel}
\end{figure}

\section{Resolution}

Why do our simulations successfully reproduce the TF relation?
 Tests[18] show that low resolution and possibly lack of feedback
 cause baryons to become too centrally concentrated due to angular
 momentum transfer from the disk to the halo (due to a grainy
 potential causing scattering of disk particles by two--body
 encounters) (Fig.4).  A galaxy too centrally concentrated would shift
 to the right of the observed TF plotted in Fig.3~[3].  As our
 simulations have a force resolution smaller than 25\% of the disk
 scale lengths and use several hundreds of thousands particles within
 the virial radius they give robust and likely converging results at
 scales corresponding to the typical scale length of stellar disks
 (1-5 kpc). Feedback also helped to prevent gas to accumulate and form
 stars at the center. A comparison between our standard MW1 run and
 its super high resolution version at z=0.5 clearly shows that as
 resolution is increased the peak V$_c$ (linked to the underlying
 density profile) and the bulge component both decrease significantly
 (Fig.4). At even lower resolutions an artificially dominant and
 concentrated bulge would have caused V$_{rot}$(3.5~R$_d$) to be
 severely overestimated, shifting simulations to the right of real
 data. {\it This test suggest that about several 10$^5$ particles and
 force resolution $<$ 1~kpc are needed to simulate the dynamics of
 galactic disks. As many as 10$^6$ DM particles and $<$~0.4
 kpc force resolution might be needed to study the formation of
 galactic bulges}.

\acknowledgements{FG would like to thank Chris Brook, Victor
 Debattista, Elena D'Onghia, Adriano Fontana, George Lake, Anatoly
 Klypin and Matthias Steinmetz for helpful and sometimes intense
 conversations.  FG was a Brooks Fellow when this project started. FG
 was supported in part by NSF grant AST-0098557 and the Spitzer Space
 Telescope Theor.  Res. Prog., provided by NASA through a contract
 issued by JPL.  We thank A.Brooks, N.Katz, L.Mayer, A.Seth,
 O.Valenzuela \& B.Willman for allowing us to show results in advance
 of publication.}

\smallskip
\noindent {\large{\bf References}}


\noindent {\footnotesize 
[1]Katz, N.\ 1992, \apj, 391, 502 
[2]Thacker, R.~J., \&  Couchman, H.~M.~P.\ 2001, \apjl, 555, L17 
[3]{Binney} J., {Gerhard} O., {Silk} J., 2001, \mnras, 321, 471,
[4]Benson et al.  2002, MNRAS, 333,177,
[5]{Navarro} J.~F., {Steinmetz} M., 2000, \apj, 538, 477 
[6]{Moore} B., {Ghigna} S., {Governato} F., {Lake} G., {Quinn} T., {Stadel} J., {Tozzi} P., 1999, \apjl, 524, L19 
[7]MacArthur,  L.~A., et al. 2004, \apjs, 152, 175 
[8]Silva, L. et al. 1998 \mnras, 509, 103
[9]~Wadsley J.~W., Stadel J., Quinn T., 2004, New Astron., 9, 137
[10]{Haardt} F.,  {Madau} P.,  1996, \apj, 461, 20
[11]{Larson}, R.~B. 1969, \mnras, 145, 405
[12]{Kennicutt} R.~C.,  1998, \apj, 498, 541
[13]{Li} Y.,  {Mac Low} M.-M.,    {Klesse} R.~S.,  2005, \apj, 626, 823
[14]Miller G.E, Scalo, J.M. 1979, ApJS, 41, 513
[15]Kroupa, P. 2002, Science, 295, 82
[16]{Raiteri} C.~M.,{Villata} M.,{Navarro} J.~F.,  1996, \aap, 315, 105
[17]{McKee}, C.~F. and {Ostriker}, J.~P., 1977, 218, 148
[18]Governato, F., et al.\ 2004, \apj, 607, 688  
[19]Giovanelli, R. et al. 1997, \aj, 113, 53  
[20]McGaugh, S.S. astro-ph/0506750, in press.
\vfill 
\end{document}